\documentclass[journal]{IEEEtran}

  \usepackage[dvips]{graphicx}
\graphicspath{{figures/}}
\usepackage[cmex10]{amsmath}
\usepackage{algorithmic}
\usepackage{amsthm}
\usepackage{verbatim, amssymb}
\usepackage{bm}
\usepackage{array}
\usepackage{eqparbox}
\usepackage{txfonts}
\usepackage{mathrsfs}
\usepackage{lipsum}
\usepackage{cite}
\usepackage{multibib}
\usepackage{mathrsfs}
\usepackage{pst-circ}

\begin{document}

\title{On Effects of Imperfect Channel State Information on Null Space Based Cognitive MIMO Communication}
\author{Shabnam~Sodagari, shabnam@ieee.org}

\maketitle

\begin{abstract}
In cognitive radio networks, when secondary users transmit in the null space of their interference channel with primary user, they can avoid interference. However, performance of this scheme depends on knowledge of channel state information for secondary user to perform inverse waterfilling. We evaluate the effects of imperfect channel estimation on error rates and performance degradation of primary user and elucidate the tradeoffs, such as amount of interference and guard distance. Results show that, based on the amount of perturbation in channel matrices, performance of null space based technique can degrade to that of open loop MIMO. Outcomes presented in this paper also apply to null space based MIMO radar waveform design to avoid interference with commercial communication systems, operating in same or adjacent bands. 
\end{abstract}


\maketitle

\section{Introduction}
\label{sec:intro}

\begin{figure}
\centering
\includegraphics[width=3in]{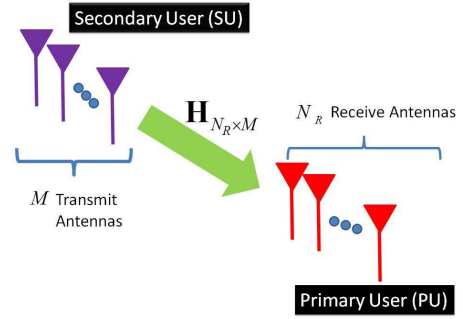}
\caption{SU transmission at null space of interference channel to avoid interference to PU}
\label{fig:fig1}
\end{figure}

We consider a coexistence scenario in which a secondary user (SU) (or a MIMO radar) is operating in close geographical proximity of a primary user (PU) and show how imperfect channel state information (CSI) in null space based SU transmission in cognitive radio networks (CRNs) can affect the performance of PU. SU has $M$ transmit antennas and PU has $N_R$ receive antennas. As in Figure~\ref{fig:fig1}, if $\textbf{x}(t)$ is the signal transmitted from SU, then the received signal at PU receiver can be written as
$\textbf{y}(t) = \textbf{H}_{N_R \times M} \textbf{x}(t) + \textbf{n}(t),$
where $\textbf{H}_{N_R \times M}$ is the interference channel matrix between SU and PU and $\textbf{n}(t)$ is the channel noise. SU maps its signals onto the null space of $\textbf{H}$. SU and PU are operating at the same frequency band, therefore we assume reciprocity of wireless channel $\textbf{H}$. Primary system can periodically inform the SU about its status, through a cognitive pilot channel (CPC) \cite{Nullspace1}. Nevertheless, in the noncooperative scenario, SU has to estimate the interference channel in order to shape its waveform in a way that does not interfere with PU \cite{Nullspace2}. 

An adaptive null space based coexistence of PU with multiple SUs for a MIMO orthogonal frequency division multiple access (OFDMA) uplink CRN is proposed in~\cite{adaptiv}. SUs transmit signals through the null-space of the channels seen between SUs and the PU base station. Lian \textit{et al.}~\cite{beamforming} also use null steering for SUs to avoid interference to PUs. Furthermore, Zarrebini-Esfahani \textit{et al.}~\cite{celledg} control interference to PUs, located outside but within the close vicinity of cognitive cell borders, by beamforming, i.e., null steering, to enhance cell-edge user coverage in cellular CRNs. 

Xiong \textit{et al.}~\cite{grassman} consider a downlink CRN scenario, with a SU base station with multiple SUs and the PU. The SU base station informs all SUs of the null space of the channel matrices between itself and the PUs. The SUs choose the beamforming vectors that are closer to the null space of the PU channels and, at the same time, maximize their projection onto the signal space of the corresponding SUs. The SUs use Grassmannian beamforming vectors from a Grassmannian codebook. Their protocol is based on the fact that only a finite number of bits are available for feeding back the CSI. Nevertheless, the amount of interference to PU is not zero, due to the quantisation of the CSI. Dai \textit{et al.}~\cite{ett-relay} perform a survey on cooperative relay channels with outdated CSI.  In addition, Zhu \textit{et al.}~\cite{interferenc} implement interference alignment in CRNs, which avoids interference to the PU by aligning the transmitted signal from each SU transmitter into the null space of the channel matrix from this transmitter to the PU. Taking into account the constraints of interference alignment in MIMO CRNs, their interference alignment algorithm is based on the minor subspace tracking that uses the fast data projection method, which does not require channel knowledge of SU. However, this scheme requires training period. 
On the other side, Hupert~\cite{opportunist} proposes to use spatial holes in multi antenna systems with a cognitive multiple access or broadcast channel. The presented methods determine a precoding matrix for each secondary transmitter as well as a post processing matrix for each secondary receiver. Based on how the precoding and the post processing matrices are determined, various methods allow for a power allocation with known waterfilling strategies in a cognitive multi user system.  

Cardoso \textit{et al.}~\cite{orthogonal} extend Vandermonde subspace frequency division multiplexing, which cancels SU interference to a primary receiver, by  exploiting the null-space of the channel from the secondary transmitter to the primary receiver. They consider how Vandermonde subspace frequency division multiplexing applies to multi-user OFDMA systems, used in Long Term Evolution (LTE). Based on weighted waterfilling, they show that interference towards multiple PUs can be cancelled, while still achieving acceptable rates for the SU. 

Yi \textit{et al.}~\cite{precoding} propose a null space-based precoding scheme for secondary transmission in a MIMO CRN under the assumption that PU uses time division duplex. To obtain the null space for precoding matrix, the SU transmitter periodically senses the transmitted signals from the PUs and estimates the corresponding covariance matrix. Next, subspace techniques are used to estimate the noise subspace of the covariance matrix, and the dimension of the noise subspace is estimated using information theory criteria. They also derive the achievable capacity of the secondary MIMO channel. Yiu \textit{et al.}~\cite{uncoordinated} consider interference cancellation and rate maximization via uncoordinated beamforming in a CRN consisting of a single PU and SU. They propose beamforming algorithms for CRNs to maximize achievable rates under the condition that the interference both at the primary and SU receivers is nullified.     
Besides, Gao \textit{et al.}~\cite{feife1} propose a SU transmission strategy consisting of three stages of environment learning, channel training, and data transmission. In the environment learning stage, the SU transceivers both listen to the PU transmission and apply blind algorithms to estimate the spaces that are orthogonal to the channels from the PU.
Assuming PU is using time division duplex transmission, SU beamforming is performed to restrict the interference to and from the PU. In the channel training stage, the SU transmitter sends training signals to SU receiver, to estimate the channel. A lower bound on the ergodic capacity for the SU in the data transmission stage is derived, taking into account imperfect estimations in learning and training stages. They also find the optimal power and time allocation for different stages.

In contrast to relevant works in the literature, here, we use techniques from perturbation theory to study the effects of null space projection method when CSI is not fully known. We show how it affects the primary system performance in terms of capacity and bit error rate (BER).

Our contributions in this paper include analysis of how errors related to interference channel estimation at the SU side can impact the performance of this technique, in terms of capacity degradation of PU and accordingly affecting interference guard distance between SU and PU. To this end, we utilize  perturbation theory concepts to evaluate differences between channel subspaces and singular values.  

The organization of this paper is as follows: Section~\ref{sec:prob} explains the problem being studied. Section~\ref{sec:imper} contains the body of our analysis. Simulation results are presented in Section~\ref{sec:simul} and Section~\ref{sec:conclus} concludes. Table~I contains definition of variables. 

\begin{table}[t]
����\begin{tabular}{|l||l|}\hline
        PU & primary user \\ \hline
        SU & secondary user \\ \hline
��������$\mathbf{H}$ & MIMO channel between SU and PU  \\ \hline
��������$\mathbf{G}$ & imperfect estimate of $\mathbf{H}$  \\ \hline
��������$\mathbf{T}$ & matrix of perturbations \\ \hline
        $N_R$ & number of PU receive antennas \\ \hline
        $M$ & number of SU transmit antennas \\ \hline
��������$\mathbf{\Theta}$ & matrix of canonical angles between two subspaces   \\ \hline
\end{tabular}
\label{tab:nota}
\caption{Notation}
\end{table}

\section{Problem Statement}
\label{sec:prob}

After the channel matrix $\textbf{H}$ is estimated, its singular value decomposition (SVD) can be obtained, as $\textbf{H} = \textbf{U}\boldsymbol\Sigma \textbf{V}^H$.
Those columns of $\textbf{V}$ corresponding to vanishing singular values in matrix $ \boldsymbol\Sigma $ span the null space of $\textbf{H}$. We denote this by $\breve{\textbf{V}}$. Note that $\textbf{V}$ is a square and unitary matrix, but $\breve{\textbf{V}}$ is not necessarily square or unitary. 
Similar to \cite{Shabnam_Globecom}, the SU signal projected onto null space of $\textbf{H}$ can be written as $\breve{\textbf{x}}(n) = \textbf{P}_{\breve{\textbf{V}}}\textbf{x}(n)$,
where
\begin{equation}\label{eqn:proj_mat}
\textbf{P}_{\breve{\textbf{V}}}= \breve{\textbf{V}}(\breve{\textbf{V}}^H \breve{\textbf{V}})^{-1}\breve{\textbf{V}}^H
\end{equation}
is the projection matrix into the null space of $\textbf{H}$, which is spanned by columns of $\breve{\textbf{V}}$. In equation (\ref{eqn:proj_mat}) $(.)^H$ denotes Hermitian of a matrix. If there are no zero singular values to perfectly define the null space, we can define a threshold greater than zero, below which we can assume the singular values to be zero and form a null space. Null space projection can be viewed as inverse waterfilling.

When SU transmission is not perfectly aligned to null space of interference channel between SU and PU, some SU transmitted power spill over to cause interference with PU. This interference depends on amount of error in estimation of channel $\mathbf{H}$ and affects capacity and BER of PU, accordingly. In this work we present an analysis of performance degradation of PU using techniques from perturbation theory. To this end, we investigate effects of imperfect CSI  on perception of interference channel singular values and subspaces, including the perturbed null space. 

\section{Solution Using Effects of Imperfect CSI}
\label{sec:imper} 

Let us denote the erroneous estimated interference channel matrix by $\mathbf{G}$. Our goal is to estimate how the null space of interference channel $\mathbf{H}$ is affected in its perturbed version $\mathbf{G}$. Nonzero singular values of $\mathbf{H}$ are square roots of nonzero eigenvalues of $\mathbf{H}\mathbf{H}^H$ or $\mathbf{H}^H\mathbf{H}$, where $(.)^{H}$ denotes Hermitian of a matrix. Estimation of effects of error on invariant subspaces of $\mathbf{H}^H \mathbf{H}$ and $\mathbf{H} \mathbf{H}^H$ provides insight into our problem. 

Before proceeding to the analysis, we introduce two common matrix norms that reduce to the Euclidean vector norm, i.e., $\|\mathbf{T}\|_2$ the spectral norm, also written $\|.\|_2$, which is defined by $\|\mathbf{T}\|_2 \overset{def}{=} \underset{\|\mathbf{x}\|_2=1}{\max} \|\mathbf{Tx}\|_2$
and the Frobenius norm, defined by 
$\|\mathbf{T}\|_\mathrm{F} \overset{def}{=}\sqrt{\underset{i,j}\sum  \mathbf{T}(i,j)^2}$.
Both of the above norms are unitarily invariant, i.e., for all unitary matrices $\mathbf{U}$ and $\mathbf{V}$ we have $\|\mathbf{U}^H\mathbf{TV}\| = \|\mathbf{T}\|$. 

\subsection{Effect of Imperfect CSI on Channel Singular Values}

When the perturbation is of order $\xi$, we have $\bm{\Sigma}+\xi \bm{\Sigma} =\mathbf{U}^H (\mathbf{H}+\xi \mathbf{H})\mathbf{V}$. Since $\mathbf{U}$ and $\mathbf{V}$ are orthogonal or unitary, they preserve norms and as a result $\left|\xi \bm\Sigma\right| = \left|\xi \mathbf{H}\right|$. In other words, perturbations of any size in knowledge of interference channel matrix cause perturbations of roughly the same size in its singular values. Perturbations and accuracy are measured relative to the norm of the matrix or, equivalently, the largest singular value. The accuracy of the smaller singular values is measured relative to the largest one.

A perturbation expansion gives an approximation of eigenvalues $\tilde{\sigma_i}$ of perturbed matrix as a function of error matrix $\mathbf{T}$ and eigenvalues  $\sigma_i$ of unperturbed matrix. In first order perturbation expansion the new eigenvalue $\tilde{\sigma}_i$ is given by $
\tilde{\sigma}_i=\sigma_i+\phi (\mathbf{T}) O(||\mathbf{T}||^2)$,
where $\phi$ is a linear function and $O(||\mathbf{T}||^2)$ means in the order of norm of channel estimation error matrix $\mathbf{T}$. For $N_R > M$. when the last $N_R-M$ columns of $\mathbf{U}$ are null vectors of $\mathbf{H}^T$, or transposition of $\mathbf{H}$, there are $N_R-M$ zero singular values. These zero singular values can affect the smaller singular values of a matrix in case of perturbation. Small singular values tend to increase with perturbation~\cite{14}. We can conclude that errors in CSI estimation, tend to confine the perceived null space and cause this space to appear smaller than its actual dimensions. This imposes restriction on SU without yielding any interference reduction on PU. In other words, the error in SU's knowledge of CSI results in useless extra provisioning on its waveform design.  One possible compensation mechanism at the SU side can be increasing the threshold on singular values of interference channel matrix $\mathbf{H}$ that define the null space.  

Although zero singular values of a rectangular matrix do not change under channel matrix estimation errors, they do affect other small singular values.   The following analysis for perturbed singular values gives more details in this regard~\cite{14,17}.
Let $P$ be the orthogonal projection onto the column space of $\mathbf{H}$. Let $P_{\perp} = \mathbf{I}-P$, where $\mathbf{I}$ is the identity matrix. 
Then, 
\begin{equation}
\label{eq:eta}
\tilde{\sigma}_i^2 = (\sigma_i + \gamma_i)^2 + \eta_i^2,
\end{equation} where $\Vert \gamma_i \Vert \leq \|P\mathbf{T}\|_2$
and $\mathrm{inf_2} (P_{\perp} \mathbf{T}) \leq \eta_i \leq \|P_{\perp} \mathbf{T}\|_2.$
In the above, the smallest singular value of the matrix $P_{\perp} \mathbf{T}$ is denoted by  $\mathrm{inf}_2 (P_{\perp} \mathbf{T})$. 
For $\sigma_i =0$, equation~(\ref{eq:eta}) gives $\tilde{\sigma_i}^2 = \gamma_i^2+\eta_i^2$. As the number of receive antennas at PU, or $N_R$, grows, $\gamma_i^2$ will on the average be of order unity, while $\eta_i^2$ will be of order $N_R$. Thus, instead of a zero singular value, we will find a nonzero singular value that tends to grow as $\sqrt{N_R}$. This shows small singular values tend to increase under perturbation proportional to $\sqrt{N_R}$.
This can result in extra interference on primary system, especially when the null space of interference channel matrix is determined by setting some positive non-zero threshold on singular values. In the case of CSI error, the previously small singular values related to correct CSI, which were used to constitute the null space, now grow to become likely greater than the threshold. This leads in the SU to mistakenly ignore some part of the null space and accordingly impose more interference on neighboring PUs. In sum, in the presence of error, the singular values corresponding to zero singular values in the unperturbed matrix will be larger than error. In particular, if the ratio of correct value to error is near $\sqrt{N_R}$, there will be error in estimating the rank. As $\sigma_i$ grows, the term $\eta_i$ becomes negligible, and the expression~(\ref{eq:eta}) becomes $\tilde{\sigma_i} \cong \sigma_i + \gamma_i$. In this case there is no upward bias in the perturbation of $\sigma_i$.

\subsection {Bounds for Singular Values with CSI Error}

Thoerems of Weyl~\cite{weyl} and Mirsky~\cite{mirsk} specify the basic bounds for the singular values of perturbed interference channel matrix. 
\newtheorem{Theorem}{Theorem}
\begin{Theorem}\textbf{(Weyl)}
$|\tilde{\sigma}_i-\sigma_i| \leq \left\|\mathbf{T}\right \|_2, i= 1, \dots, M,$
\end{Theorem}
where $M$ is the number of SU transmit antennas. We denote the capacity of PU with exact CSI by $C$ and the capacity with erroneous CSI by $\tilde{C}$.
Since the interference level from SU to PU (or from cognitive MIMO radar to the neighboring communication system~\cite{Shabnam_Globecom}) is related to knowledge of singular values of interference channel, we can conclude that, according to Weyl's theorem, the capacity degradation of PU, as a result of erroneous CSI, is loosely upper bounded by the following relation
\begin{equation}
|C-\tilde{C}| < \left\|\mathbf{T}\right \|_2, i= 1, \dots, M.
\end{equation}
In other words, since capacity of PU is directly proportional to accurate knowledge of null space by SU, which is related to knowledge of eigenvalues, any error in estimation of eigenvalues results in some SU power to spill over from null space and interfere with PU. This interference degrades capacity of PU. As a result, an error in knowledge of singular values, which according to Weyl theorem is upper bounded by $\left\|\mathbf{T}\right \|_2$ guides us to amount of capacity degradation of PU.  

\begin{Theorem}\textbf{(Mirsky)}
\begin{equation}
\label{eq:mirsk}
\sqrt{\sum_i (\tilde{\sigma}_i-\sigma_i)^2} \leq \left\|\mathbf{T}\right \|_\mathrm{F}
\end{equation}
\end{Theorem}
Mirsky's theorem includes Weyl's theorem as a special case. Furthermore, Mirsky's theorem holds for an arbitrary unitarily invariant norm.
There is no restriction on the size of the error and the theorems are true for any $\mathbf{T}$~\cite{stew90}. Weyl's theorem states that the singular values of a matrix are perfectly conditioned in that no singular value can move more than the norm of the perturbations. If we divide both sides of equation (\ref{eq:mirsk}) by $\sqrt{M}$, we see that Mirsky's theorem means the root mean square of the errors in the singular values is bounded by the root mean square of the singular values of the error. Although Mirsky's theorem is less precise than Weyl's theorem, it is usually more useful, since the Frobenius norm is easy to calculate. If a singular value is small compared with norm of $\mathbf{T}$, it may be entirely wiped out. Therefore, though singular values are perfectly conditioned, there is no guarantee that they are determined with high accuracy.
Mirsky's theorem tends to overestimate the variation of the singular values. The perturbation expansion can be used to give some insight into the amount of this overestimation. As a special case, let us suppose there are $N_R-M$ zero singular values for interference channel matrix $\mathbf{H}$ with dimension $N_R \times M$. This can happen when $N_R > M$ and the interference channel matrix is full rank. Elements of $\mathbf{T}$ are independent random variables with mean zero and standard deviation $\rho$. Then, if the singular values of $\mathbf{H}$ are simple (not repeated) and second order terms are ignored, the perturbation in the $i$th singular value is $\mathbf{U}^T_i \mathbf{T}\mathbf{V}_i$, where $\mathbf{U}_i$ and $\mathbf{V}_i$ are the corresponding left and right singular vectors. Thus, with $M$ SU transmit antennas, the expected value of the sum of squares of the errors in the singular value is $\mathbb{E}\left\lbrack \sum_{i} \left(\mathbf{U}_i^T \mathbf{T} \mathbf{V}_i\right)^2\right\rbrack = M\rho^2,$
while the expected value of the square of the Frobenius norm of $\mathbf{T}$ is 
$\mathbb{E} (\|\mathbf{T}\|_\mathrm{F}^2)= N_R M\rho ^2$. Here, $\mathbb{E}$ denotes expectation. 
Thus, Mirsky's theorem tends to overestimate the root mean square error in the singular value by a factor of square root of PU receive antennas $\sqrt {N_R}$. However, when one singular value is repeated multiple times, Mirsky's bound can be sharp. 

\subsection{Effect of Imperfect CSI on Subspaces}
\label{subsec:subspac}

Singular vectors corresponding to close singular values are extremely sensitive to even small errors. There exist MIMO interference channel matrices for which arbitrarily small errors completely change the singular vectors. This can particularly affect $\breve{\mathbf{V}}$ in equation~(\ref{eqn:proj_mat}).

A useful approach is to compute bounds for the subspace spanned by the singular vectors, called singular subspace. This is due to the fact that in case of errors in CSI estimation individual singular vectors corresponding to a cluster of singular values are unstable.

Our goal is to compare singular subspaces spanned by singular vectors for both cases of perfect CSI and imperfect CSI. Therefore, we need to define distance metrics between the two subspaces, such as canonical angles~\cite{stew90}. For two one dimensional subspaces $\mathcal{W}$ and $\mathcal{Z}$, the angle between them is given by
$(\mathcal{W},\mathcal{Z}) = \arccos |\mathbf{w}^T \mathbf{z}|,$
where $\mathbf{w}$ and $\mathbf{z}$ are vectors of norm one spanning $\mathcal{W}$ and $\mathcal{Z}$.
To generalize this to $k$ dimensional subspaces, let $\mathcal{W}$ and $\mathcal{Z}$ be subspaces of dimension $k$. Let $\mathbf{W}$ and $\mathbf{Z}$ be orthonormal bases for $\mathcal{W}$ and $\mathcal{Z}$, with
$c_1 \geq \dots \geq c_k$
being the singular values of $\mathbf{W}^T\mathbf{Z}$. The numbers $\theta_i = \arccos c_i$ are called the canonical angles between $\mathcal{W}$ and $\mathcal{Z}$. In other words, $\mathcal{W}$ and $\mathcal{Z}$ are close if the largest canonical angle is small. Next, we explain how canonical angles are connected to projections. 
Let $P_{\mathcal{W}}$ and $P_{\mathcal{Z}}$ be the orthogonal projections onto $\mathcal{W}$ and $\mathcal{Z}$. We can take $\|P_{\mathcal{W}}-P_{\mathcal{Z}}\|$ as another measure of the distance between $\mathcal{W}$ and $\mathcal{Z}$, for if $\mathcal{W}=\mathcal{Z}$, then  $P_{\mathcal{W}}=P_{\mathcal{Z}}$. We note that the two measures, i.e., canonical angles and projections are the same and they go to zero at the same rate, because according to~\cite{stew90}~$\|P_{\mathcal{W}}-P_{\mathcal{Z}}\|_\mathrm{F} = \sqrt{2} \|\sin\Theta\|_\mathrm{F}$, where $\Theta$ is the matrix of canonical angles between the two subspaces.  
The capacity degradation of PU is proportional to the amount of spill over power from SU to PU, due to null space misalignment, which is proportional to the difference between the two pertinent null spaces
\begin{equation}
C-\tilde{C} \propto \|\sin\Theta\|_\mathrm{F}. 
\end{equation}

\subsection{Bound for Null Space of Interference Channel with Imperfect CSI}

Wedin theorem~\cite{wedin} derives perturbation bounds for singular subspaces. This theorem provides a single bound for both the right and left singular subspaces corresponding to a set of singular spaces. 
Let 
\begin{eqnarray}
(\mathbf{U}_1 \mathbf{U}_2 \mathbf{U}_3)^H \mathbf{H} (\mathbf{V}_1 \mathbf{V}_2) = \begin{bmatrix}  \bm\Sigma_1 & 0 \\ 0 &\bm\Sigma_2 \\ 0 &0 \end{bmatrix},
\end{eqnarray}
be a singular value decomposition of $\mathbf{H}$, in which the singular values are not necessarily in descending order. The singular subspaces are the column spaces of $\mathbf{U}_1$ and $\mathbf{V}_1$ and are bounded according to Wedin Theorem. The perturbed subspaces will be the columns spaces of $\tilde{\mathbf{U}}_1$ and $\tilde{\mathbf{V}}_1$ in the decomposition of $\mathbf{G}$
\begin{eqnarray}
\left(\tilde{\mathbf{U}}_1 \tilde{\mathbf{U}}_2 \tilde{\mathbf{U}}_3\right) ^H  \mathbf{G} \left(\tilde{\mathbf{V}}_1 \tilde{\mathbf{V}}_2\right) = \begin {bmatrix} \tilde{\bm\Sigma}_1 & 0 \\ 0& \tilde{\bm\Sigma}_2 \\ 0 &0 \end{bmatrix} 
\end{eqnarray}
Let $\Phi$ be a matrix of canonical angles between subspaces spanned by $\mathbf{U}_1$ and $\tilde{\mathbf{U}}_1$, and let $\Theta$ be the matrix of canonical angles between subspaces spanned by $\mathbf{V}_1$ and $\tilde{\mathbf{V}}_1$. Wedin theorem derives bounds on $\Phi$ and $\Theta$. The bounds are not directly expressed in terms of $\mathbf{T}$, but in terms of the residuals $
\mathbf{R} = \mathbf{H} \tilde{\mathbf{V}}_1 - \tilde{\mathbf{U}}_1 \tilde{\Sigma}_1~~\mathrm{and}~~
\mathbf{S} = \mathbf{H}^H \tilde{\mathbf{U}}_1 - \tilde{\mathbf{V}}_1 \tilde{\Sigma}_1.$
Note that if $\mathbf{T}$ is zero, then $\mathbf{R}$ and $\mathbf{S}$ are zero. More generally, $
\|\mathbf{R}\| \leq \|(\mathbf{G} - \mathbf{T}) \tilde{\mathbf{V}}_1 - \tilde{\mathbf{U}}_1 \tilde{\bm{\Sigma}}_1\| \leq  \|\mathbf{T}\tilde{\mathbf{V}}_1\| \leq \|\mathbf{T}\|$
with a similar bound for $\mathbf{S}$. 
\begin{Theorem}\textbf{(Wedin)}~\cite{stew90}
\label{wedin}
If there is a $\delta >0$ such that 
\begin{eqnarray}
\label{eq:cond1}
\min\vert \sigma (\tilde{\Sigma}_1) -\sigma (\Sigma_2)\vert \geq \delta~~\mathrm{and}~~
\label{eq:cond2}
\min \sigma (\tilde{\Sigma}_1) \geq \delta
\end{eqnarray}
then 
\begin{equation}
\label{eq:wedn_bound}
\sqrt{\|\sin \Phi\|^2_\mathrm{F} + \|\sin \Theta\|^2_\mathrm{F}}\leq \frac{\sqrt{\|\mathbf{R}\|^2_\mathrm{F}+\|\mathbf{S}\|^2_\mathrm{F}}}{\delta}.
\end{equation}
\end{Theorem}
The conditions~(\ref{eq:cond1}) are separation conditions. The first expresses that the singular values in $\bm\Sigma_1$ are separated from those in $\bm\Sigma_2$. In fact, the separation is between $\tilde{\Sigma}_1$ and $\Sigma_2$. Nevertheless, when $\mathbf{T}$ is small in comparison with $\delta$, then according to Weyl's theorem, the two expressions are equivalent. The second condition, which is a necessary condition, indicates that the singular values in $\Sigma_1$ are separated from the ghost singular values. The same bound as in (\ref{eq:wedn_bound}) holds for the space spanned by $\mathbf{V}_1$. 

\subsection{Distance between Null Spaces as Trigonometric Function of their Angle}

There exists a singular value decomposition of $\mathbf{H}$~\cite{wedin}:
\begin{eqnarray}
\label{eq:h}
\mathbf{H} = \mathbf{U} \bm{\Sigma} \mathbf{V}^H = \mathbf{U}_1 \bm{\Sigma}_1 \mathbf{V}_1^H + \mathbf{U}_0 \bm{\Sigma}_0 \mathbf{V}^H_0,
\end{eqnarray}
where 
\begin{align}
&\mathbf{V}_1=[\mathbf{v}_1,\dots,\mathbf{v}_r],~\mathbf{V}_0=[\mathbf{v}_{r+1}, \dots, \mathbf{v}_p], \\ \nonumber &\mathbf{V} = (\mathbf{V}_1, \mathbf{V}_0)\\
&\mathbf{U}_1=[\mathbf{u}_1, \dots, \mathbf{u}_r],~\mathbf{U}_0=[\mathbf{u}_{r+1}, \dots, \mathbf{u}_p],\\ \nonumber &\mathbf{U} = (\mathbf{U}_1, \mathbf{U}_0)\\
&\bm{\Sigma}_1 = \mathrm{diag} (\sigma_1, \dots, \sigma_r),~\bm{\Sigma}_0 = \mathrm{diag} (\sigma_{r+1}, \dots, \sigma_p),\\&\bm{\Sigma}= \mathrm{diag} (\sigma_1, \dots, \sigma_p)
\end{align}

The rank of $\mathbf{H}$ is $p$ and $r \leq p$. Here, $\mathbf{V}_1, \mathbf{V}_0, \mathbf{V}$ and $\mathbf{U}_1, \mathbf{U}_0, \mathbf{U}$ are assumed to be partial isometries satisfying
\begin{align}
&\mathbf{V}^H \mathbf{V} = \mathbf{U}^H \mathbf{U} =\mathbf{I}_p,~\mathbf{V}_1^H \mathbf{V}_1 =\mathbf{U}_1^H \mathbf{U}_1 =\mathbf{I}_r,\\ 
\nonumber &\mathbf{V}_0^H \mathbf{V}_0 =\mathbf{U}_0^H \mathbf{U}_0 =\mathbf{I}_{p-r}
\end{align}

For the perturbation of $\mathbf{H}$ or $\mathbf{G} = \mathbf{H}+\mathbf{T}$ the SVD can be written as
\begin{eqnarray}
\label{eq:g}
\mathbf{G}_1= \mathbf{U}(\mathbf{G})_1 \bm{\Sigma}(\mathbf{G})_1 \mathbf{V}(\mathbf{G})_1^H~\mathrm{and}~\\ \nonumber
\mathbf{G}_0= \mathbf{U}(\mathbf{G})_0 \bm{\Sigma}(\mathbf{G})_0 \mathbf{V}(\mathbf{G})_0^H
\end{eqnarray}
Denote the null space and range of $\mathbf{H}$ by $N(\mathbf{H})$ and $R(\mathbf{H})$, respectively. From equations (\ref{eq:h}) and (\ref{eq:g})   
\begin{eqnarray}
R(\mathbf{H}_1)=N(\mathbf{H}_1^H)^\perp ~~~~\mathrm{and}~~~~N(\mathbf{H}_1)=R(\mathbf{H}_1^H)^\perp,
\end{eqnarray}
where $\perp$ denotes orthogonal complement. 

For any unitary invariant norm, the angle between two subspaces $\mathcal{W}$ and $\mathcal{Z}$ is defined as~\cite{wedin}: $\left\|\sin\theta(\mathcal{W},\mathcal{Z})\right\|=\left\|(\mathbf{I}-\mathbf{P}_{\mathcal{Z}}) \mathbf{P}_{\mathcal{W}}\right\|$.
Upper bounds for $\left\|\sin\theta\left(R(\mathbf{G}_1^H),R(\mathbf{H}_1^H)\right)\right\|$ have been derived in \cite{wedin}. In this regard, residuals that replace $\mathbf{T}$ need to be defined. Let $\mathbf{Y}_1=[\mathbf{y}_1,\dots,\mathbf{y}_r]$ contain orthonormal vectors $\mathbf{y}_i,~i=1,\dots,r$ spanning the subspace $R(\mathbf{G}_1)$ and let $\mathbf{X}_1=[\mathbf{x}_1,\dots,\mathbf{x}_r]$ contain orthonormal vectors spanning $R(\mathbf{G}_1^H)$. Then,
\begin{eqnarray}
\mathbf{Y}_1^H \mathbf{Y} = \mathbf{I_r};~~\mathbf{Y}_1^H \mathbf{Y} = P_{R(\mathbf{G}_1)};\\ \nonumber~~\mathbf{X}_1^H \mathbf{X} = \mathbf{I_r};~~\mathbf{X}_1^H \mathbf{X} = P_{R(\mathbf{G}_1^H)} 
\end{eqnarray}
With choice of $\mathbf{X}_1 = \mathbf{V}_1$ and $\mathbf{Y}_1= \mathbf{U}_1$, define $\mathbf{D}_1 = \mathbf{Y}_1^H \mathbf{G} \mathbf{X}_1 = \Sigma_1(\mathbf{G}_1)$~\cite{wedin}. By defining the residuals as follows
\begin{eqnarray}
\label{eq:residu}
 R_{11}=\mathbf{H} \mathbf{X}_1-\mathbf{Y}_1 \mathbf{D}_1~~\mathrm{and~~} R_{21}=\mathbf{H}^H \mathbf{Y}_1- \mathbf{X}_1 \mathbf{D}_1^H 
\end{eqnarray}
we observe that the residuals are related to $\mathbf{T}$ by the following 
\begin{align}
&R_{11}= \mathbf{H}\mathbf{X}_1 - \mathbf{Y}_1 \mathbf{D}_1 \\ \nonumber &=(\mathbf{G}-\mathbf{T})\mathbf{X}_1 -\mathbf{Y}_1 (\mathbf{Y}_1^H \mathbf{G} \mathbf{X}_1)=-\mathbf{T}\mathbf{X}_1 \mathrm{~and~}\\\nonumber 
&R_{12}=(\mathbf{G}^H - \mathbf{T}^H) \mathbf{Y}_1-\mathbf{X}_1 (\mathbf{X}_1^H \mathbf{G}^H \mathbf{Y}_1) = -\mathbf{T}^H \mathbf{Y}_1
\end{align}
Similarly, define $\mathbf{Y}_0$ and $\mathbf{X}_0$ corresponding to $R(\mathbf{G}_0)$ and $R(\mathbf{G}_0^H)$. Then, $
P_{R(\mathbf{G}_0^H)} =\mathbf{X}_0 \mathbf{X}_0^H;~\mathbf{I} =\mathbf{X}_0^H \mathbf{X}_0;~
P_{R(\mathbf{G}_0)} = \mathbf{Y}_0 \mathbf{Y}_0^H;~\mathbf{I} = \mathbf{Y}_0^H \mathbf{Y}_0.$
Define $\mathbf{D}_0=\mathbf{Y}_0^H \mathbf{G} \mathbf{X}_0$. The corresponding residuals are $
R_{01}= \mathbf{H}\mathbf{X}_0 - \mathbf{Y}_0 \mathbf{D}_0~~\mathrm{and~~} 
R_{02}=\mathbf{H}^H \mathbf{Y}_0 -\mathbf{X}_0 \mathbf{D}_0^H.$
Using above notation, we can now proceed to the $\sin\theta$ theorem for singular value decomposition~\cite{wedin}.
\begin{Theorem}
\label{Theorem1} 
Assume there exists an $\alpha \geq 0$ and a $\delta > 0 $, such that 
$\sigma_{\mathrm{min}}(\mathbf{G}_1) \geq \alpha + \delta  ~\mathrm{and}~\sigma_{\mathrm{max}} (\mathbf{H}_0) \leq \alpha.$
With $R_{11}$ and $R_{12}$ defined by equation~(\ref{eq:residu}), set
$\epsilon = \max (||R_{11}||,||R_{12}||).$
Then, for every unitary invariant norm, 
\begin{eqnarray}
\left\|\sin\theta(R(\mathbf{H}_1),R(\mathbf{G}_1))\right\| \leq  \frac{\epsilon}{\delta}~\\ \nonumber \mathrm{and~}
\left\|\sin\theta(R(\mathbf{H}_1^H),R(\mathbf{G}_1^H))\right\| \leq  \frac{\epsilon}{\delta}.
\end{eqnarray}
\end{Theorem}
As a result, when $\mathbf{H}$ is perturbed $P_{R(\mathbf{H}_0)}$ and $P_{R(\mathbf{H}_0^H)}$ are influenced not only by $P_{R(\mathbf{H}_1)}$ and $P_{R(\mathbf{H}_1^H)}$, but also by $P_{R(\mathbf{H})^\perp}$ and $P_{N(\mathbf{H})}$. 

In this case, denote energy per bit of the primary system by $E_p$ and energy per bit of secondary system by $E_s$. Let $N_0$ denote thermal noise at PU receiver. With perfect null space projection, noting that probability of error is directly related to Q function of signal to interference plus noise ratio, error probability at PU receiver or $P_e$ is proportional to  
\begin{eqnarray}
\label{eq:pe}
P_e \propto Q\left(\sqrt{\frac{E_p}{N_0}}\right)
\end{eqnarray}
Imperfect null space projection aggravates probability of error at PU, owing to added interference from SU, proportional to spill over power from SU to PU. This interference is denoted by $\frac{E_s}{d^{-\varsigma}} \|\sin \Theta\|$, wherein $d$ is the distance between SU transmitter and PU receiver and $\varsigma$ is an attenuation factor. Note that when the two subspaces are perfectly aligned, i.e., $\|\sin \Theta\| =0$, the interference amounts to zero. Therefore, 
\begin{eqnarray}
\tilde{P}_e \propto Q\left(\sqrt{\frac{E_p}{N_0+\frac{E_s}{d^{-\varsigma}} \|\sin \Theta\|}}\right). 
\end{eqnarray}
Using Theorem 4, 
\begin{eqnarray}
\sqrt{\frac{E_p}{N_0+\frac{E_s}{d^{-\varsigma}} \|\sin \Theta\|}} \geq \sqrt{\frac{E_p}{N_0+\frac{E_s}{d^{-\varsigma}} (\frac{\epsilon}{\delta})}}. 
\end{eqnarray}
In other words, since $Q$ function is monotonically decreasing, probability of error is upper bounded by 
\begin{eqnarray}
\label{eq:upbound}
\tilde{P}_e \leq Q\left(\sqrt{\frac{E_p}{N_0+\frac{E_s}{d^{-\varsigma}} (\frac{\epsilon}{\delta})}}\right).
\end{eqnarray}

With perfect null space projections, the capacity of PU is proportional to
\begin{equation}
C \propto \log \frac{E_p}{N_0} 
\end{equation} 
However, with imperfection projection, the degraded capacity of PU is inversely proportional to amount of interference from SU, i.e.,
\begin{eqnarray} 
\label{eq:cap}
\tilde{C} \propto \log \frac{E_p}{N_0+\frac{E_s}{d^{-\varsigma}} \|\sin \Theta\|}.
\end{eqnarray}
Hence, 
\begin{align}
C-\tilde{C} &\propto \log \frac{E_p}{N_0} - \log \frac{E_p}{N_0+\frac{E_s}{d^{-\varsigma}} \|\sin\Theta\|}  \\ \nonumber& =
 \log \frac{N_0+\frac{E_s}{d^{-\varsigma}}\|\sin \Theta\|}{N_0} \leq \log \frac{N_0+\frac{E_s}{d^{-\varsigma}}\frac{\epsilon}{\delta}}{N_0}
\end{align}

If $N(\mathbf{H})$ or $N(\mathbf{G})$ is nonempty, which is our desired case, we need a lower bound of $\sigma_{\mathrm{min}}(\mathbf{G}_0)$ in the theorem above to be able to estimate $||\sin\theta (R(\mathbf{G}_0^H),R(\mathbf{H}_0^H))||$. On this point, we consider an extension of original $\sin\theta$ theorem~\cite{wedin}.
\begin{Theorem}
\label{sinus_extend}
Assume there is an interval $[\beta,\alpha]$ and a $\delta \geq 0$ such that the singular values of $\mathbf{G}_0$ lie entirely in $[\beta,\alpha]$, while the singular values of $\mathbf{H}_1$ lie entirely outside of $(\beta-\delta,\alpha +\delta)$ (or such that the singular values of $\mathbf{H}_1$ lie entirely in $[\beta,\alpha]$, while those of $\mathbf{G}_0$ lie entirely outside of $(\beta-\delta,\alpha +\delta)$). Set~$\epsilon = \max (\left\|R_{01},R_{02}\right\|$.
Then, for every unitary invariant norm 
\begin{align}
\label{eq:sin_extend}
&\max (||\sin \theta(R(\mathbf{G}_0),R(\mathbf{H}_0))||, ||\sin \theta(R(\mathbf{G}_0^H),R(\mathbf{H}_0^H))||)\\ \nonumber
&\leq \frac{\epsilon}{\delta}k
\end{align} with $k=\sqrt{2}$ for spectral norm and the Euclidean matrix norm and $k \leq 2$ for all unitary invariant norms. \begin{proof} For detailed proof of Theorems 4 and 5, refer to~\cite{wedin}.
\end{proof}
\end{Theorem}

In this case, using Theorem 5 and following same line of reasoning that led to deriving equation (\ref{eq:upbound}) from Equation (\ref{eq:pe}), we obtain 
\begin{eqnarray}
\label{eq:upbound2}
\tilde{P}_e \leq Q\left(\sqrt{\frac{E_p}{N_0+\frac{E_s}{d^{-\varsigma}} (\frac{\epsilon}{\delta})k}}\right) 
\end{eqnarray}

Note that in this case the degraded capacity is proportional to
\begin{eqnarray} 
\tilde{C} \propto \log \frac{E_p}{N_0+\frac{E_s}{d^{-\varsigma}}E_s \|\sin \Theta\|} 
\end{eqnarray}
Since acording to equation (\ref{eq:sin_extend}) the difference between two subspaces is less than $\frac{\epsilon}{\delta}k$, we obtain
\begin{eqnarray} 
\log \frac{E_p}{N_0+\frac{E_s}{d^{-\varsigma}} \|\sin \Theta\|} \geq \log \frac{E_p}{N_0+\frac{E_s}{d^{-\varsigma}}(\frac{\epsilon}{\delta})k}
\end{eqnarray}
Therefore, a lower bound for degraded capacity is 
\begin{eqnarray} 
\tilde{C} \geq \log \frac{E_p}{N_0+\frac{E_s}{d^{-\varsigma}} (\frac{\epsilon}{\delta})k}.
\end{eqnarray}

\section{Simulation Results}
\label{sec:simul}

\begin{figure}[h!]
\centering
\includegraphics[width=3.2in]{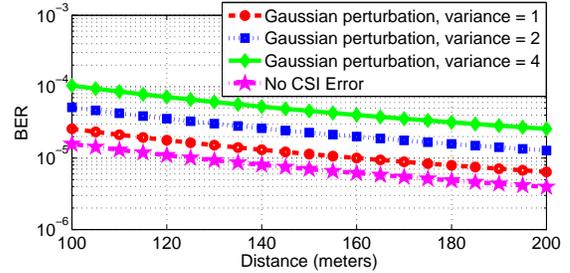}
\caption{BER vs. PU and SU distance for Gaussian perturbations}
\label{fig:bergaus}
\end{figure}

\begin{figure}[tbh]
\centering
\includegraphics[width=3.2in]{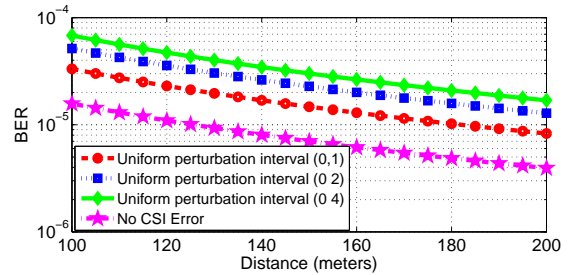}
\caption{BER vs. PU and SU distance for uniform perturbations}
\label{fig:berunif}
\end{figure}

\begin{figure}[tbh]
\centering
\includegraphics[width=3.7in]{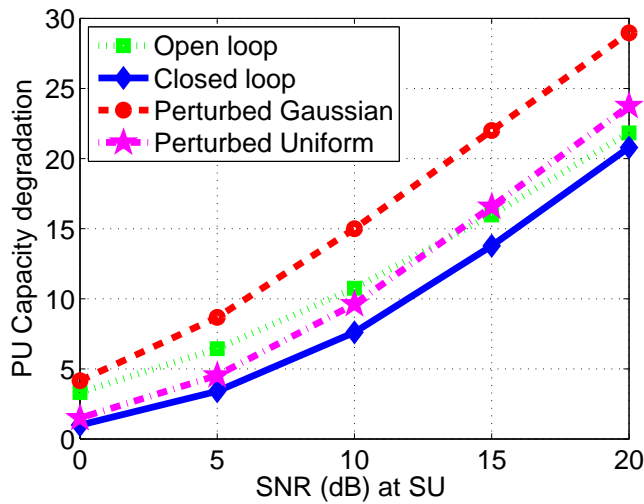}
\caption{Capacity degradation of PU vs. SU transmit power}
\label{fig:loop} 
\end{figure}

\begin{figure}[tb]
\centering
\includegraphics[width=3.7in]{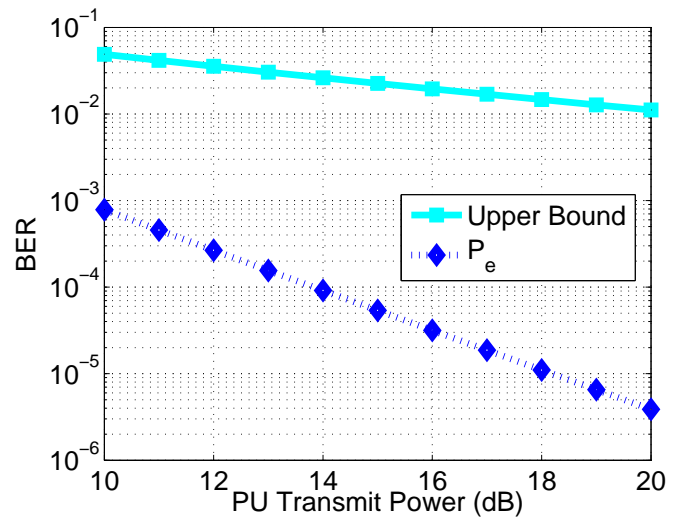}
\caption{BER and BER upper bound vs. PU transmit power}
\label{fig:bound} 
\end{figure}

Figure~\ref{fig:bergaus} compares BER imposed upon PU as a result of perturbations in CSI, for Gaussian errors with zero mean and different variances, with the perfect CSI BER case. Note the nonzero BER for the case of perfect CSI knowledge is due to choice of a nonzero threshold for eigenvalues to define the null space. As expected, as the distance between PU receiver and SU transmitter increases, BER on PU decreases due to reduced amount of interference, owing to path loss. Figure~\ref{fig:berunif} contains same information for uniformly distributed perturbations in CSI. All figures are related to binary phase shift keying (BPSK) $2\times 2$ MIMO first order diversity Rician fading channels. The Rician specular to diffuse energy ratio is set to 3. Results indicate that errors in CSI estimation can affect the null space transmission and introduce almost one order of magnitude higher BERs. 
Figure~\ref{fig:loop} shows errors in CSI in null space based coexistence of PU and SU even have more severe effects on PU than open loop MIMO, especially when SU transmit power increases. Figure~\ref{fig:bound} shows BER vs. transmit power of SU in dB and upper bound derived for BER in equation~(\ref{eq:upbound2}). As Figure~\ref{fig:bound} demonstrates, the upper bound always holds for BER of PU system as a result of power spilling from SU due to imperfect null space alignment. 

\section{Conclusion}
\label{sec:conclus}

We analyzed impacts of imperfect channel matrix estimations on null space based coexistence of PU and SU in cognitive MIMO communication systems, which interrupts inverse waterfilling (null space projection) on account of displaced singular values and subspaces. We benchmarked the tradeoffs in degradation of PU performance vs. interference guard distance between the two systems and SU transmit powers, for different error types. 

\bibliographystyle{IEEEtrans}
\bibliography{IEEEabrv,Trans1}

\end{document}